\documentclass[onecolumn,reprint,aps,prb,floatfix,superscriptaddress]{revtex4}%
\usepackage{amssymb,amsbsy,graphicx}
\usepackage[latin1]{inputenc}
\begin{document}
\title{Analysing surface structures on (Ga,Mn)As  by Atomic Force Microscopy}

\author{S. Piano}
\email{samanta.piano@nottingham.ac.uk}
\affiliation{School of Physics and Astronomy, University of Nottingham, University Park, Nottingham, NG7 2RD, United Kingdom}
\author{A. W. Rushforth}
\affiliation{School of Physics and Astronomy, University of Nottingham, University Park, Nottingham, NG7 2RD, United Kingdom}
\author{K. W. Edmonds}
\affiliation{School of Physics and Astronomy, University of Nottingham, University Park, Nottingham, NG7 2RD, United Kingdom}
\author{R. P. Campion}
\affiliation{School of Physics and Astronomy, University of Nottingham, University Park, Nottingham, NG7 2RD, United Kingdom}
\author{G. Adesso}
\affiliation{School of Mathematical Sciences, University of Nottingham, University Park, Nottingham, NG7 2RD, United Kingdom}
\author{B. L. Gallagher}
\affiliation{School of Physics and Astronomy, University of Nottingham, University Park, Nottingham, NG7 2RD, United Kingdom}

\begin{abstract}
Using atomic force microscopy, we have studied the surface structures of high quality molecular beam epitaxy grown (Ga,Mn)As compound. Several samples with different thickness and Mn concentration, as well as a few (Ga,Mn)(As,P) samples have been investigated. All these samples have shown the presence of periodic ripples aligned along the $[1\overline{1}0]$ direction. From a detailed Fourier analysis we have estimated the period ($\sim 50 {\rm nm}$) and the amplitude of these structures.
\end{abstract}

\maketitle
\section{Introduction}

The modern life always demands new technologies that should be fast, mobile and at low cost. It is then necessary to miniaturise current devices, and a question arises whether we need to adapt the existing technologies or build one ``ex novo''. While the limits of present-day technologies have been exposed, as miniaturisation of current computer architectures towards nanoscale will be hindered by quantum mechanical effects, on the other hand quantum mechanics itself can also help us by offering currently unused electrons properties to be efficiently exploited. The conventional classical devices rely on the charge of the electrons to produce energy, encode and manipulate information, but recently new and useful devices are being developed that utilise the spin of the electrons as a carrier and processor of information, bearing the promise to create a new generation of devices which will be smaller, more versatile and more robust than those currently made by semiconductors.
 In this context,   the family of (III,Mn)V ferromagnetic semiconductors represent prominent candidates for the spintronics industry. In particular the (Ga,Mn)As compound  has attracted much attention for its potential applications in non-volatile memories, spin-based optoelectronics and quantum computation.\cite{Ohno,Loss}. Ferromagnetic semiconductors are interesting also from a fundamental physics
side, in particular to understand the nature of the magnetic interactions that underlie the ferromagnetism and to
investigate the microscopic origin of the magnetic anisotropy in these compounds \cite{Tomas, Tomas_2}.
It has been known that ferromagnetic (Ga,Mn)As films are characterised by a
substantial magnetic anisotropy \cite{Anisotropy, Anisotropy_2} (consisting of a cubic and a uniaxial components),
which has been in general related to strain, hole concentration
and  temperature.\cite{Dietl, Abolfath,Welp,Sawicki}.  Several
attempts have been performed to tune the magnetic anisotropy by
designing 1D nano-objects using lithographic methods but the
possibility of growing self organised ordered 1D structures has
remained unexplored so far and triggered the work presented in this paper. Very recently, we applied atomic force microscopy and grazing incidence X-ray diffraction measurements to reveal the presence of ripples on the surface of (Ga,Mn)As layers grown on GaAs(001) substrates and buffer layers \cite{APL}. A connection between the surface anisotropy that characterises the distribution of the ripples and the uniaxial magnetic anisotropy has been suggested, deserving further investigation.

In this paper we report a detailed study of the structural
surface properties of (Ga,Mn)As by means of atomic force microscopy (AFM).
This technique is regarded as a very useful tool to
investigate the surface of non-conductive samples with a very high
resolution at {\rm nm} scale \cite{Binning, review}. We have analysed a variety of (Ga,Mn)As samples with thickness ranging from $5$ to $25 {\rm nm}$, observing self-organised periodic ripples aligned along the $[1\overline{1}0]$ crystallographic direction on the surface of all the measured samples. By using a Fourier power spectral density (PSD) analysis \cite{Fang}, we have obtained a quantitative model of the periodicity of the surface ripples for each sample, calculating their amplitude and effective period. The amplitude is related to the root mean square roughness, which is around $(0.38 \pm 0.10) {\rm nm}$ for all the observed samples, while the effective period, which provides an  estimate of the width of each ripple, stays around $(53 \pm 12) {\rm nm}$. We have further analyzed a handful of (Ga,Mn)(As,P) samples finding no significant change in their structural properties compared with the case of P-free compounds. We argue that the combination of
the AFM data and the PSD analysis provides a complete and quantitative description of the surface of the (Ga,Mn)As compound.

\section{Experimental data and analysis}

\begin{figure}[t!]
\centering \includegraphics [width=14cm]{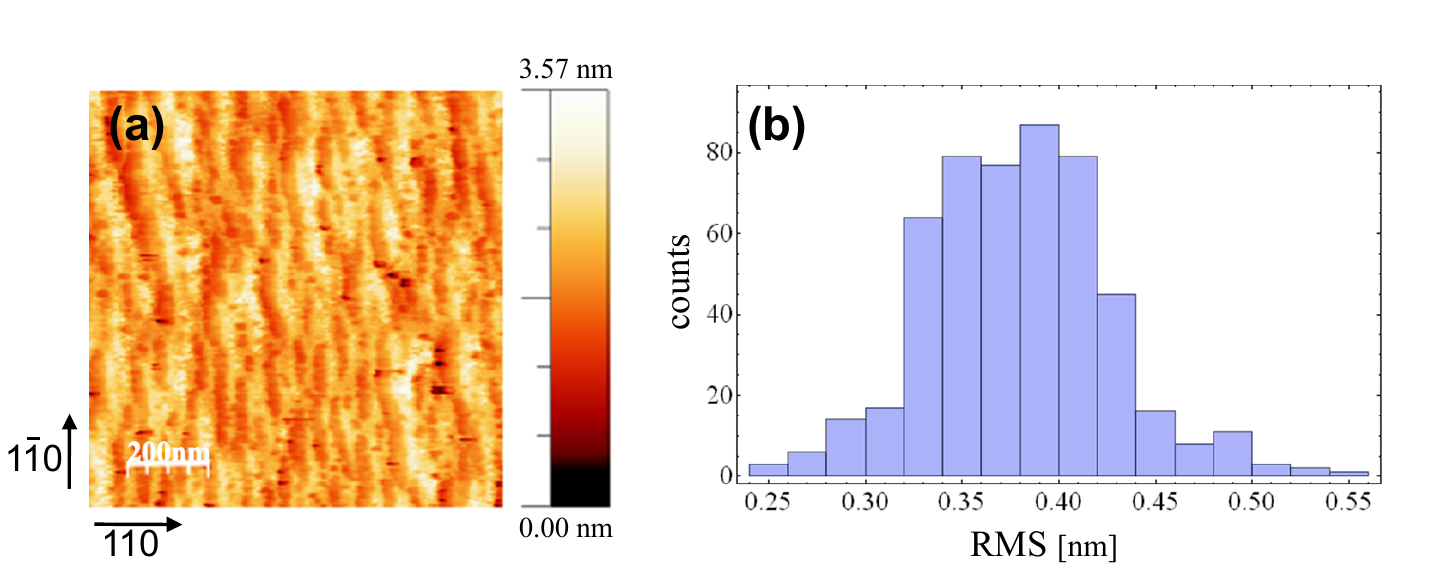}
\caption{(a) AFM image of (Ga,Mn)As sample Mn351, $5{\rm nm}$, 2\% Mn; self-organised ripples grow along the $[1\overline{1}0]$ direction. (b) Histogram of the RMS roughness distribution across the horizontal scan lines for the same sample; its global RMS is $(0.38 \pm 0.04) {\rm nm}$.} \label{image1}
\end{figure}

\begin{figure*}[tb]
\includegraphics [width=16cm]{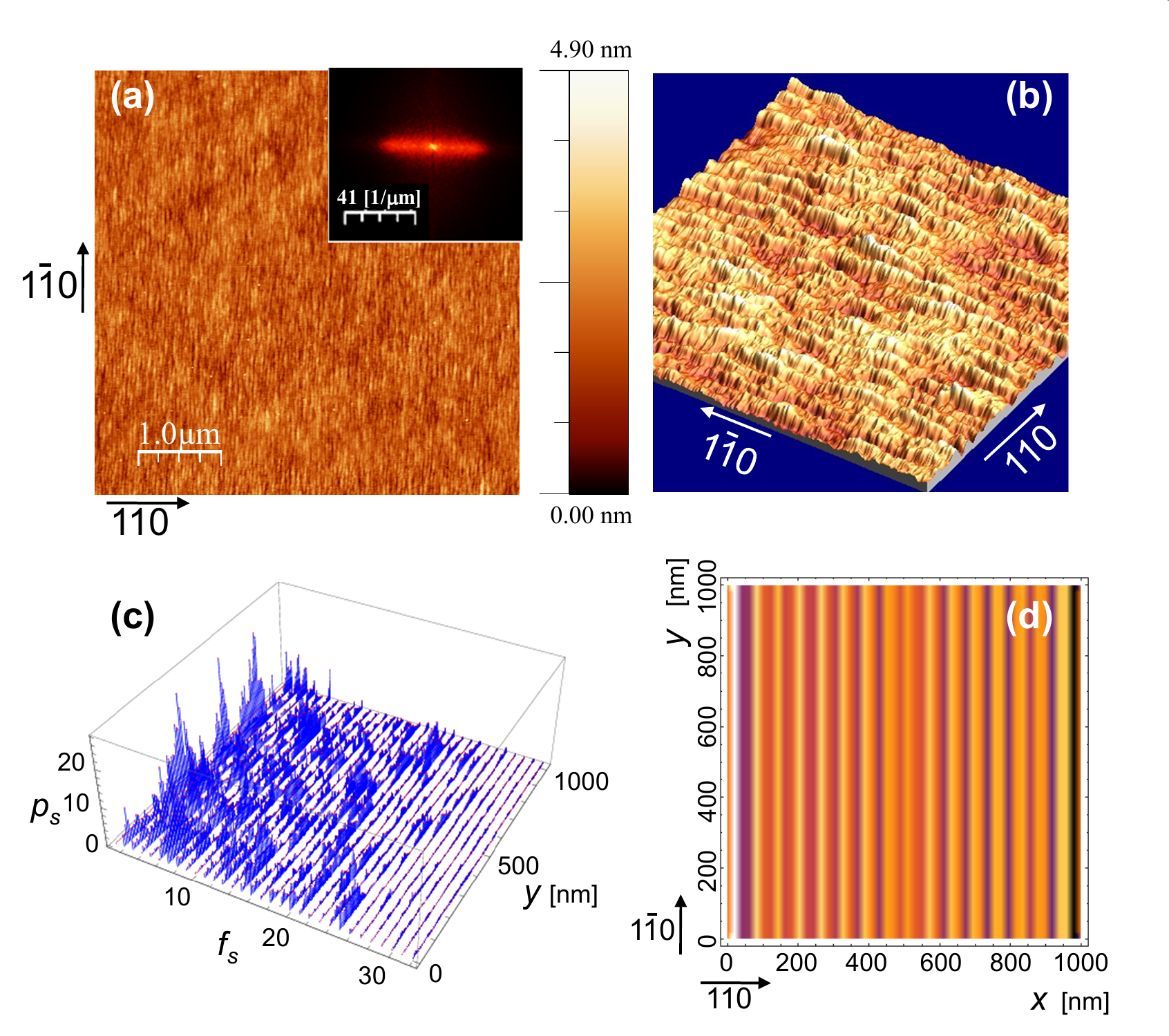}
\caption{AFM data and Fourier PSD analysis on the (Ga,Mn)As sample Mn381. (a) AFM image of a $5 \times 5 \mu{\rm m}^2$ area of the sample, with neat evidence of the ripples along the $[1\overline{1}0]$ direction; the inset shows the 2D PSD spectrum, clearly signaling the anisotropy of the surface. (b) 3D AFM image of an $1 \times 1 \mu{\rm m}^2$ area of the same sample. (c) Line-by-line PSD showing the Fourier pattern  along each horizontal scan line, for the image in (b). Reconstructed image data from a filtered Fourier transform where only a narrow window of harmonics around the effective mean period $\tau \pm 2\delta\tau$ is retained, for the sample in (b).} \label{image2}
\end{figure*}

(Ga,Mn)As  films with different thickness,  $5 {\rm nm}$, $7 {\rm nm}$ and $25
{\rm nm}$ and Mn concentrations, 6\% and 12\%, and (Ga,Mn)(As,P) films with different P concentrations, 3\%, 6\% and 9\%,  have been deposited on
GaAs(100) substrates by low temperature molecular beam epitaxy,
the details are described elsewhere. \cite{preparation,prep2} AFM images
were obtained by using an Asylum Research MFP-3D atomic force
microscope and Asylum research silicon probes have been used as tips.
Before performing  measurements the samples have been
cleaned with a 1:3 $HCl$:$H_2O$ to remove
the superficial oxide layer. Measurements on uncleaned samples
have shown the same surface structures but with additional surface
contamination indicating that the cleaning was not modifying the
surface morphology. The majority of the images were taken on $1\times 1 \mu{\rm m}^2$ areas
along two different crystallographic direction, $[1\overline{1}0]$
and $[110]$. In Figs. \ref{image1}(a) and \ref{image2}(b,c)  we show some typical images for two different samples.
 We can see, clearly, that some ripples grow aligned along the $[1\overline{1}0]$ direction. Similar surface structures have been observed in all the measured samples, including the phosphorate ones.

To have an estimation of the periodicity of the features emerging from the AFM images we have used the Fourier analysis. The inset of Fig. \ref{image2}(a) shows a typical 2D PSD spectrum. One can clearly see the surface anisotropy from the shape of the graph, showing that periodicity across the horizontal direction is strongly enhanced compared to the vertical one (for an isotropic image, the 2D PSD spectrum should accumulate along the 45$^\circ$ diagonal). In order to properly take into account the observed anisotropy, and to achieve a more accurate quantitative information about the structural periodic features, we decompose our images into horizontal scan lines and apply Fourier analysis to each line, combining the data at the end.

The heights along a horizontal scan line can be collected in a vector $z(y)=\{z_r\}$, where the index $r$ runs from $1$ to the number $n$ of points (e.g. $n=1024$).
The fast Fourier transform  can be computed as
\begin{equation}
\phi_s=\frac{1}{\sqrt{n}} \sum_{r=1} ^n z_r e^{2 \pi i (r-1)(s-1)/n}
\end{equation}
here $\phi_s$ is the Fourier amplitude corresponding to a (dimensionless) frequency $f_s=s-1$. Each harmonics, corresponding to a periodic function with period $T_s=L/f_s$, is given by: $z_r=\frac{1}{\sqrt{n}} \sum_{s=1} ^n \phi_s e^{-2 \pi i (r-1)(s-1)/n}$, where $L$  is the window length in ${\rm nm}$ (e.g. $L=1000$).
The PSD is defined as $p_s=|\phi_s|^2$ [see Fig. \ref{image2}(c)].

For each horizontal scan line, we have calculated the RMS roughness as standard deviation of the $z(y)$ height data.  The RMS can be equivalently computed from the total PSD and it is equal (up to a factor
$\sqrt{2}$) to the amplitude ($A$) of the oscillations of the
periodic structures:
\begin{equation}
A=\sqrt{2} {\rm RMS}=\frac{\sqrt{2 \sum_{s=1}^n p_s}}{n-1}
\end{equation}
An histogram showing an exemplary distribution of the RMS roughness across the horizontal scan lines is depicted in Fig. \ref{image1}(b).
For each image we have calculated the ``effective'' global RMS (and its error) as
the mean (and standard deviation) of the roughnesses per each horizontal scan line. The  values of the global RMS vary between
$0.28 {\rm nm}$ for a sample $25 {\rm nm}$ thick with $6\%$ Mn and $0.48 {\rm nm}$
for a sample $7 {\rm nm}$ thick with $6\%$ Mn.

Concerning the periodicity of the structures, we define the mean frequency per line, $\overline{f}(y)$,  as an average of the frequencies of all Fourier harmonics (along a given horizontal scan line) weighted with the corresponding PSDs:
\begin{equation}
\overline{f}(y)=\frac{\sum_{s=1}^{n/2}(s-1)p_s}{\sum_{s=1}^{n/2}p_s}
\end{equation}
Then, the global ``effective'' mean frequency ($\overline{f}$), characterising the oscillating ripple pattern in the whole image, is given by the mean of the $\overline{f}(y)$'s of each line $y$ along all horizontal scan lines, and the associated error $\delta\overline{f}$ is the standard deviation of the $\overline{f}(y)$'s.
From the mean frequency $\overline{f}$ and the window length $L$ we obtain the ``effective'' period $\tau$ of the surface structures,
\begin{equation}
 \tau=\frac{L}{\overline{f}}.
\end{equation}
Such an effective period $\tau$ is expected to give a quantitative estimate of the mean width of a single  ripple observed in the AFM
images.  The uncertainty $\delta\tau$ on the effective period is calculated accordingly, from $\delta\overline{f}$.
The regularity of the patterns and the {\it bona fide} characterisation of the morphograms by means of an effective Fourier component is further confirmed by observing that the relative error $\delta\tau/\tau$ on the effective period stays below 20\% for each image.
Furthermore, by filtering only those $\lesssim 10$ harmonics (out of a total of $256$ or $512$) that lie within two standard deviations around the effective mean period (or, equivalently, in a frequency window $\overline{f} \pm 2 \delta\overline{f}$) one can reconstruct, via Fourier transform of the PSD, a theoretical periodic image that displays an overlap with the original experimental image of at least 50\%, for all the reported samples [see Fig. \ref{image2}(d)].
The effective period $\tau$ for all the $20$ different  samples is about $(53 \pm 12) {\rm nm}$.

\begin{table}[t!]
\begin{tabular}{llllll}
\hline \hline
   & Sample & Thickness& RMS (nm) &  $\tau$ (nm)\\
   \hline \hline

 1 & Mn 351 & 5 nm, 2\%  &	0.38  $\pm$ 0.04 &	 \quad 42.3 $\pm$ 4.2\\
  2 & Mn 352 &	5 nm, 6\% & 0.35 $\pm$ 0.07 	  & \quad49.7 $\pm$ 12.7 \\
  3 & Mn 556	& 5 nm, 6\%  &	0.38 $\pm$ 0.05  & \quad 45.14 $\pm$ 11.5 \\
  4 & Mn 536 & 5 nm, 6\%  &	0.31 $\pm$ 0.04	   & \quad45.5 $\pm$ 11.3\\
   5 & Mn 396 & 7 nm, 6\%  & 0.48 $\pm$ 0.06    & \quad 47.7 $\pm$ 8.2\\
  6 & Mn 381 & 7 nm, 6\%	&   0.38 $\pm$ 0.04	  & \quad 41.9 $\pm$ 6.4 \\
    7 & Mn 555 	& 7 nm, 6\%  &	0.3 $\pm$ 0.1	    & \quad 47.7 $\pm$ 6.3\\
    8 & Mn 535	& 7 nm, 6\%  &	0.34 $\pm$ 0.04	   & \quad 45.4 $\pm$ 10.3\\
     9 & Mn 394 & 7 nm, 6\%  & 	0.35 $\pm$ 0.05	   & \quad 59.8 $\pm$ 13.5\\

  \hline

  10 & Mn 437 & 25 nm, 12\% & 0.36 $\pm$ 0.06  & \quad  51.5 $\pm$ 8.6\\
   11 & Mn 438 & 25 nm, 12\% & 0.37 $\pm$ 0.07 &   \quad 61.7 $\pm$ 11.4\\
   12 & Mn 439 & 25 nm, 12\%  &	0.41 $\pm$ 0.05	  &  \quad  44.4 $\pm$ 5.1\\
  \hline

  13 & Mn 467 & 25 nm 6\% & 0.29 $\pm$ 0.07  &   \quad 57.4 $\pm$ 8.6\\
      14 & Mn 554	& 25 nm, 6\%  &	0.35 $\pm$ 0.04	&   \quad 56.0 $\pm$ 16.9\\
  15 & Mn 499 & 25 nm, 6\%  &	0.31 $\pm$ 0.05	   &   \quad 47.8 $\pm$ 11.8\\
     16 & Mn 330  &	25 nm, 6 \% & 0.31 $\pm$ 0.09    & \quad 65.9 $\pm$ 11.3\\
        17 & Mn 490 & 25 nm, 6\%  &	0.28 $\pm$ 0.03	    &  \quad 52.5 $\pm$ 0.03\\
        \hline
         18 & Mn 491 & 25nm, 6\% P 6\% Mn &	0.28 $\pm$ 0.04	 &  \quad  51.5 $\pm$ 8.2\\
         19 & Mn 492	& 25nm, 3\% P 6\% Mn &	0.38 $\pm$ 0.08	 &  \quad  54.2 $\pm$ 12.7\\
         20 & Mn 498	& 25nm, 9\% P 6\% Mn &	0.37 $\pm$ 0.08	  &  \quad  52.61 $\pm$ 10.4\\

 \hline \hline
\end{tabular}
\caption{Summary of the properties of (Ga,Mn)As samples analysed with the AFM. Notation:
RMS is the global roughness; $\tau$ is the ``effective'' period of the ripples.}\label{table1}
\end{table}

In table \ref{table1} we summarise the main properties of the measured samples: thickness, \% Mn, effective period, and global RMS roughness.

\begin{table}[t!]
\begin{tabular}{lll}
  \hline \hline
  Sample Mn 381 & Scan area ($\mu{\rm m}^2$)& $\tau$ ({\rm nm})\\
  \hline\hline
  1 & 1x1 & 37.44 \\
  2 & 5x5 & 45.16 \\
  3 & 1x1 & 42.77 \\
  4 & 1x1 & 36.6 \\
  5 & 1x1 & 39.2 \\
  6 & 5x5 & 35 \\
  7 & 10x10 & 44.86 \\
  8 & 2x2 & 40.5\\
  \hline \hline
\end{tabular}
\caption{Effective period $\tau$ calculated from the Fourier analysis on different areas of the sample Mn381, showing consistent values of the periodicity of the ripples.}\label{table2}
\end{table}

To confirm that the observed structures extend to the whole surface area of the grown (Ga,Mn)As compounds, we have performed additional measurements on  one chosen sample, Mn381, selected for the particularly neat quality and visibility of its periodic surface ripples (see Fig.~\ref{image2}).  We have acquired many AFM images on different areas of the sample with different size. The effective period $\tau$ for several $1 \times 1 \mu{\rm m}^2$, $2\times 2 \mu{\rm m}^2$, $5 \times 5 \mu{\rm m}^2$ and $10 \times 10 \mu{\rm m}^2$ AFM images is found to be stable at about $(40 \pm 5) {\rm nm}$. See Table \ref{table2} for a summary of the thickness and the Fourier periodicity figures of the ripples for different AFM images on this sample.

\section{Conclusions}

In conclusion we have studied the surface properties of (Ga,Mn)As compound with the AFM. From the analysis of different samples we can estimate a roughness of about $(0.38 \pm 0.10) {\rm nm}$. All of the measured samples show the formation of self-organised ripples along $[1\overline{1}0]$ crystallographic direction with an effective period of about $(53 \pm 12) {\rm nm}$, as estimated from a Fourier analysis.
Completely analogous features  (qualitatively and quantitatively) have been revealed for (Ga,Mn)(As,P) samples with varying P concentration. While it is known that epitaxially grown (Ga,Mn)(As,P) samples show magnetic properties comparable with P-free (Ga,Mn)As, the P layer does in general alter the strain state from compressive to tensile, inducing a modification of the magnetic anisotropy.\cite{prep2,muu} Nonetheless,  we have noticed no significant difference in the structural surface properties, compared with the samples without P. A more extended case study with more finely tuned P concentration is certainly needed to support this conclusion, and will be the subject of further investigation.

The presence of anisotropic surface structures in (Ga,Mn)As has been recently confirmed by grazing incidence X-ray diffraction measurements\cite{APL}, resulting in an estimate of the periodicity which is consistent with the findings reported in this work. An interesting future direction will be to investigate the mechanism at the heart of the formation of the observed ripples, in particular the role of lattice strain, and the correlation with the uniaxial magnetic anisotropy of (Ga,Mn)As. This is expected to shed new light on the magnetic properties of this material and in general of the family of (III,Mn)V ferromagnetic semiconductors. Developing a deeper understanding of how the ferromagnetic interactions between the local Mn moments, mediated by the itinerant holes, give rise to the observed magnetic properties, is essential in order to unleash the potential of (Ga,Mn)As to realise manyfold applications for the spintronics industry.


\acknowledgments{Funding from the EU (Grant Nos.
NAMASTE-214499 and SemiSpinNano-237375) is acknowledged.}

\end{document}